\documentclass[a4paper,twocolumn,amsmath,amssymb,aps,preprintnumbers,superscriptaddress,showpacs,longbibliography]{revtex4-1}

\pdfoutput=1 

\usepackage[latin9]{inputenc}
\usepackage{textcomp}
\usepackage{amsmath}
\usepackage{amssymb}
\usepackage{graphicx}

\usepackage{dcolumn}
\usepackage{bm}
\usepackage{epsfig}
\usepackage[usenames]{color}
\usepackage{xcolor}

\definecolor{pinegreen}{rgb}{0.0, 0.47, 0.44}
\definecolor{persiangreen}{rgb}{0.0, 0.65, 0.58}
\definecolor{pakistangreen}{rgb}{0.0, 0.4, 0.0}
\definecolor{mossgreen}{rgb}{0.68, 0.87, 0.68}

\newif\ifcom
\newif\ifdel

\comtrue                    
\deltrue                      

\begin{document}
\title{Impedance spectroscopy of ferroelectrics: The domain wall pinning element}

\author{M.~Becker}
\email{maximilian.becker@nmi.de}
\affiliation{%
NMI Natural and Medical Sciences Institute at the University of T\"ubingen,
Markwiesenstr.~55,
72770 Reutlingen, Germany}
\affiliation{%
Physikalisches Institut, Center for Quantum Science (CQ) and LISA$^+$,
University of T\"ubingen,
Auf der Morgenstelle 14,
72076 T\"ubingen, Germany}

\author{C.~J.~Burkhardt}
\affiliation{%
NMI Natural and Medical Sciences Institute at the University of T\"ubingen,
Markwiesenstr.~55,
72770 Reutlingen, Germany}

\author{R.~Kleiner}
\affiliation{%
Physikalisches Institut, Center for Quantum Science (CQ) and LISA$^+$,
University of T\"ubingen,
Auf der Morgenstelle 14,
72076 T\"ubingen, Germany}

\author{D.~Koelle}
\affiliation{%
Physikalisches Institut, Center for Quantum Science (CQ) and LISA$^+$,
University of T\"ubingen,
Auf der Morgenstelle 14,
72076 T\"ubingen, Germany}

\date{\today}

\begin{abstract} 
We introduce an equivalent-circuit element based on the theory of interface pinning in random systems, to analyze the contribution of domain wall motion below the coercive field to the impedance of a ferroelectric, as a function of amplitude $E_0$ and frequency $f$ of an applied ac electric field.
We investigate capacitor stacks, containing ferroelectric 0.5(Ba$_{0.7}$Ca$_{0.3}$)TiO$_{3}$--0.5Ba(Zr$_{0.2}$Ti$_{0.8}$)O$_{3}$ (BCZT) thin films, epitaxially grown by pulsed laser deposition on Nb-doped SrTiO$_3$ single crystal substrates and covered with Au electrodes.
Impedance spectra from $f=10$\,Hz to 1\,MHz were collected at different $E_0$. 
Deconvolution of the spectra is achieved by fitting the measured impedance with an equivalent-circuit model of the capacitor stacks, and we extract the domain-wall-motion induced amplitude- and frequency-dependent dielectric response of the BCZT films from the obtained fit parameters.
From an extended Rayleigh analysis, we quantify the coupling strength between dielectric nonlinearity and dielectric dispersion in the BCZT films and identify different domain-wall-motion regimes.
Finally, we construct a schematic diagram of the different domain-wall-motion regimes and discuss the corresponding domain-wall dynamics.
\end{abstract} 

\maketitle

\section{Introduction} 
\label{sec:Introduction}

Domain wall motion in ferroelectric materials does not only influence their ferroelectric, but also their dielectric and piezoelectric properties \cite{Damjanovic98}.
In general, the complex dynamics of domain wall displacement in ferroelectrics results in a dependence of their dielectric $(\varepsilon_{ij})$ and piezoelectric  $(d_{ijk})$ tensor components on the amplitude $E_0$ and frequency $f$ of an applied ac electric field \cite{Damjanovic98,Damjanovic05}.

Regarding perovskite-type ferroelectrics, field and frequency dependent dielectric and piezoelectric response has been mostly investigated in ceramics and thin films based on PbZr$_{x}$Ti$_{1-x}$O$_{3}$ (PZT) \cite{Damjanovic98,Damjanovic05,Bassiri-Gharb05,Bassiri-Gharb07}.
It was found, that the field dependence of permittivity and piezoelectric coefficients at subswitching fields can often be described by the linear Rayleigh law, which results in a nonlinear and hysteretic dielectric polarization and piezolectric strain. 
Based on this, the linear field dependence of permittivity is also referred to as dielectric nonlinearity \cite{Bassiri-Gharb07}.
Moreover, the coupling of dielectric nonlinearity and frequency dispersion has been studied \cite{Damjanovic05}.
However, a unified theory which describes the field and frequency behavior of material properties, e.g. the complex relative permittivity $\varepsilon_r(E_0,f)$ of ferroelectrics, with the same set of equations is not available.
Thus, the measurement of $\varepsilon_r(E_0,f)$  is crucial, since the permittivity is directly related to the microstructure of the material and essential for device modelling \cite{Damjanovic98, McKinstry04}.
This is especially true for thin film samples of new and thus less investigated lead-free ferroelectrics such as K$_{0.5}$Na$_{0.5}$NbO$_{3}$ (KNN) and 0.5(Ba$_{0.7}$Ca$_{0.3}$)TiO$_{3}$--0.5Ba(Zr$_{0.2}$Ti$_{0.8}$)O$_{3}$ (BCZT) \cite{Saito04,Liu09}.

Recently, a new method for the analysis of the field dependence of complex relative permittivity $\varepsilon_r(E_0)$ (i.e. Rayleigh analysis) was introduced \cite{Schenk18}. 
This method is based on impedance spectroscopy and equivalent-circuit fitting, 
which allows one to discriminate different contributions to the permittivity, originating from electronically different components (electrodes, eletrode/ferroelectric interfaces and ferroelectric) of the analyzed thin film devices.

In thin film capacitor stacks, with a ferroelecric placed between two metallic electrodes, the dielectric properties can exhibit strong frequency dispersion depending on the resistivity of both, the ferroelectric film and the electrode layers \cite{Tyunina06}, thus masking the true dielectric dispersion of the ferroelectric caused by pinning of domain walls on randomly distributed defects occuring in the ferroelectric layer inside the heterostructure \cite{Nattermann1990,Fedorenko04}.
Moreover, the interfaces to the electrodes play a more crucial role in thin films as compared to thick ceramic samples \cite{Schenk18,Schmidt2007}, which can lead to extrinsic dielectric relaxation processes \cite{Schmidt2012}, additionally masking the contributions from the ferroelectric.

The superiority of the Rayleigh analysis based on impedance spectroscopy over the conventional approach (i.e. without equivalent-circuit fitting) was demonstrated in Ref.~\onlinecite{Schenk18} for the measurement of $\varepsilon_r(E_0)$ for a ferroelectric Si:HfO$_{2}$ thin film.
However, the frequency dispersion $\varepsilon_r(f)$ is not accessible with the method described in Ref.~\onlinecite{Schenk18}.

In this work, we introduce an equivalent-circuit element with impedance  $Z_{\rm DW}$ which is based on the theory of interface pinning in random systems \cite{Nattermann1990} and models the frequency-dependent impedance response of a ferroelectric including the domain-wall-motion induced contribution.
Moreover, we report on a  method to extract the domain-wall-motion induced amplitude- and frequency-dependent dielectric response $\varepsilon_r(E_0,f)$ in thin film ferroelectrics, embedded in a thin film capacitor stack, from impedance spectroscopy measurements at different $E_0$ and equivalent-circuit fits with the introduced element $Z_{\rm DW}$.
Finally, we present an extended Rayleigh analysis which allows one to quantify the coupling strength between dielectric nonlinearity and frequency dispersion and the identification of different domain-wall-motion regimes.
Our approach will be demonstrated by applying it to thin film capacitor stacks containing the lead-free ferroelectric BCZT.

\section{Theory}
\label{sec:Theory}

\subsection{Rayleigh behavior and interface pinning in random systems}
\label{subsec:Rayleigh-behavior}

Domain wall dynamics below the coercive field $E_{\rm C}$ of a ferroelectric can be considered as a superposition of center of mass motion and relaxational motion of internal modes, i.e. domain wall segments \cite{Fedorenko04}. 
From theory it is known, that the interaction of domain walls with pinning centers, that form a random energy landscape, gives rise to the so called Rayleigh behavior and the corresponding Rayleigh law \cite{Kronmueller70,Boser87}.
Rayleigh behavior originates from the irreversible center of mass motion of domain walls and results in a linear dependence of permittivity and piezoelectric coefficients on the field amplitude $E_0$ above a specific threshold field $E_{\rm T}$ and below the coercive field $E_{\rm C}$, which was also observed experimentally on ceramic and thin film samples \cite{Damjanovic97,Taylor97,Hall99,Bassiri-Gharb07}.
As a consequence of Rayleigh behavior, the dielectric polarization and piezoelectric strain responses of ferroelectrics become nonlinear in $E_0$ and hysteretic even at subswitching fields $E_{{\rm T}}<E_{0}<E_{{\rm C}}$.
In the following, we will solely discuss dielectric properties and disregard piezoelectric properties.

The Rayleigh law for the real and imaginary parts of the complex relative permittivity $\varepsilon_r=\varepsilon_r'-i\varepsilon_r''$ is given by \cite{Hall99,Schenk18}
\begin{eqnarray} 
\varepsilon_{r}' (E_{0})& = & \varepsilon_{r,\text{in}}'+\alpha' E_{0}\quad ,\nonumber \\
\varepsilon_{r}'' (E_{0})& = & \varepsilon_{r,\text{in}}''+\alpha'' E_{0} \quad .
\label{eq:epsilon(E_0)}
\end{eqnarray} 
Here $\varepsilon_{r,\text{in}}'$ and $\varepsilon_{r,\text{in}}''$ are real and imaginary parts of the reversible (initial) Rayleigh parameter $\varepsilon_{r,\text{in}}$ due to lattice and reversible domain wall contributions,
$\alpha'$ denotes the irreversible Rayleigh parameter (Rayleigh constant) and $\alpha''=\frac{4}{3\pi}\alpha'$.

In the following, we will denote a general linear field dependene of $\varepsilon_{r}' (E_{0})$ and $\varepsilon_{r}'' (E_{0})$ with positive slope as "Rayleigh-like behavior".
In that sense, the Rayleigh law is a special case of Rayleigh-like behavior with the additional requirement $\alpha''=\frac{4}{3\pi}\alpha'$.

The irreversible domain wall contribution to the real part of the permittivity is given by $\alpha' E_0$.
It turns out, that $\alpha'$ is a direct and quantitative measure for the mobility of domain walls, and it is dependent on the concentration of pinning centers, film thickness and the density of domain walls \cite{Bassiri-Gharb07}.
However, the Rayleigh law Eq.~(\ref{eq:epsilon(E_0)}) is rate-independent and no frequency dependence of material properties can be derived from it \cite{Bassiri-Gharb05,Damjanovic05,Bassiri-Gharb07}.

Theory predicts for domain wall pinning processes in random systems a logarithmic dependence of permittivity on frequency $f$, which is due to the relaxational contribution of internal modes \cite{Nattermann1990,Fedorenko04}
\begin{eqnarray} 
\varepsilon_{r}' (f)& \propto & \left[\ln\left(\frac{1}{2 \pi f t_0}\right)\right]^{\Theta} \quad ,\nonumber \\
\varepsilon_{r}'' (f)& \propto & \left[\ln\left(\frac{1}{2 \pi f t_0}\right)\right]^{\Theta-1} \quad .
\label{eq:epsilon(f)_allg}
\end{eqnarray} 
Here, $\Theta$ is related to the roughness of the domain wall and $t_0$ denotes a microscopic time constant, which is related to the rate of transitions of domain wall segments between metastable states \cite{Nattermann1990,Fedorenko04}.

In this pinned regime, the center of mass of the domain wall  is captured in a potential valley, while domain wall segments can still jump between different metastable states with close energies.
The potential valley of a random energy landscape can be described by \cite{Fedorenko04}
\begin{eqnarray} 
U(\bar{r}) & = & \frac{C_n}{2 \tau_1} \bar{r}^2 + \int_{0}^{\bar{r}} dr' \bar{g}(r') \quad ,
\label{eq:U_random}
\end{eqnarray} 
with a constant $C_n$, the transverse spatial coordinate  $\bar{r}$,  a single relaxation time $\tau_1$  and the random pinning force $g$. 
The first term in Eq.~(\ref{eq:U_random}) describes the average (parabolic) shape of the well and the second term describes spatial fluctuations due to the residual part of the random pinning force $g$. This term is important, because it results in a distribution of relaxation times $\Psi_{1}(\tau)$ instead of the single relaxation time $\tau_1$ \cite{Fedorenko04}.
As a consequence, the dielectric response of a ferroelectric due to domain wall pinning in a random energy landscape is not expected to exhibit a Debye-like dielectric relaxation with a single relaxation time and hence one dominant relaxation frequency.

In many ferroelectrics - including relaxors - it was found that $\Theta \sim 1$, which corresponds to a  frequency-independent imaginary part of permittivity, and which is valid for systems with a broad distribution $\Psi_{1}(\tau)$ of relaxation times \cite{Damjanovic1997,Mueller04,Damjanovic05,Bassiri-Gharb05,Bassiri-Gharb07,Yang19}.
In this case, the logarithmic frequency dependence of the complex relative permittivity can be well approximated by the expression \cite{Mueller04,Damjanovic05,Bassiri-Gharb05,Yang19} 
\begin{eqnarray} 
\varepsilon_{r}' (f)& = & \varepsilon_{r,0}'+ \Delta \varepsilon_{r}' \ln\left(\frac{1\,{\rm Hz}}{2 \pi f  }\right) \quad ,\nonumber \\
\varepsilon_{r}'' (f)& = & \varepsilon_{r,0}'' \quad ,
\label{eq:epsilon(f)}
\end{eqnarray} 
where $\varepsilon_{r,0}'$ and $\varepsilon_{r,0}''$ denote static contributions to the permittivity, and  $\Delta \varepsilon_{r}'$ is the logarithmic dispersion strength.
We note that in  Eq.~(\ref{eq:epsilon(f)}) we fixed the value of $t_0$ to 1s, which is also the common choice in the literature \cite{Damjanovic1997,Mueller04,Bassiri-Gharb07,Griggio12,Yang19}.
We will use this convention throughout the remaining part of this paper.

It has been shown for systems with Rayleigh-like behavior and logarithmic frequency dispersion, that the frequency dependence of $\varepsilon_{r}'$ in  Eq.~(\ref{eq:epsilon(f)}) is due to the logarithmic frequency dependence of both reversible  ($\varepsilon_{r,\text{in}}'$) and irreversible ($\alpha'$) Rayleigh parameters defined in the Rayleigh law  Eq.~(\ref{eq:epsilon(E_0)}), which can then be expressed as \cite{Damjanovic1997}
\begin{eqnarray} 
 \varepsilon_{r,\text{in}}'(f)& = & \varepsilon_{r,\text{in},0}'+ \Delta  \varepsilon_{r,\text{in}}' \ln\left(\frac{1\,{\rm Hz}}{2 \pi f}\right) \quad ,\nonumber \\
\alpha'(f)& = & \alpha_{0}'+ \Delta \alpha'  \ln\left(\frac{1\,{\rm Hz}}{2 \pi f}\right) \quad .
\label{eq:RayleighParameter}
\end{eqnarray} 
The physical interpretation of the coefficients $\Delta  \varepsilon_{r,\text{in}}'$ and $\Delta \alpha'$  in Eq.~(\ref{eq:RayleighParameter}) will be discussed later in Sec. \ref{sec:Results}.
As a consequence, the Rayleigh relation for the real part of relative permittivity  $\varepsilon_{r}'$ in  Eq.~(\ref{eq:epsilon(E_0)}), can be written with frequency-dependent Rayleigh parameters given by  Eq.~(\ref{eq:RayleighParameter}), as
\begin{eqnarray} 
\varepsilon_{r}' (E_{0},f)& = & \varepsilon_{r,\text{in}}'(f)+\alpha'(f) E_{0}\quad .
\label{eq:epsilon(E_0,f)}
\end{eqnarray} 

However, the Rayleigh law Eq.~(\ref{eq:epsilon(E_0)}), comprises also the Rayleigh relation for the imaginary part of relative permittivity $\varepsilon_{r}''$ and the relations are coupled by  $\alpha''=\frac{4}{3\pi}\alpha'$.
The coupling relation predicts a logarithmic frequency dependence of $\alpha''$ which results in a frequency dispersion of $\varepsilon_{r}''$, which is contradictory to the assumption of a frequency independent imaginary part given by Eq.~(\ref{eq:epsilon(f)}), which implies a constant  $\alpha''$.
Thus, the logarithmic frequency dispersion due to domain wall pinning in random systems and the Rayleigh law are only compatible at a  fixed frequency.
Taking account of a constant $\alpha''$, we can implicitly define this frequency $f \equiv f_{\rm{R}}$ by
\begin{eqnarray} 
\lim\limits_{f \rightarrow f_{\rm{R}}} \frac{\alpha'(f)}{\alpha''} & \stackrel{!}{=} &\frac{3\pi}{4} \quad.
\label{eq:fR}
\end{eqnarray} 
At $f =f_{\rm{R}}$, the Rayleigh law is fulfilled for variable field amplitudes $E_0$ within the Rayleigh region.
The frequency $f_{\rm{R}}$ will be called the "Rayleigh frequency" in the following.

The more general Rayleigh-like behavior exists in the frequency range where $\alpha'(f)$ is positive. 
At frequencies $f$ within this Rayleigh-like frequency range, the center of mass of domain walls can move irreversibly via hopping between different minima of the energy landscape, since this is a basic assumption in the derivation of the Rayleigh law.
Thus, the center of mass of the domain walls is not captured in a potential valley as in the pinned regime and we conclude that the Rayleigh-like frequency range corresponds to a regime with coexisting center of mass motion and relaxational motion of internal modes.
In analogy to Ref.~\onlinecite{Fedorenko04}, we will denote this regime as the stochastic regime of complex relative permittivity.

\subsection{Derivation of an equivalent-circuit element based on interface pinning in random systems}
\label{subsec:IS}

In an impedance spectroscopy experiment, the complex impedance $Z(f)=Z'(f)-iZ''(f)$ is measured as a function of frequency $f$, typically in the range of  $f=10 - 10^6$ Hz.
The complex impedance is related to the complex relative permittivity by \cite{Gerhardt94}
\begin{eqnarray} 
Z (f)& = & \frac{1}{i 2 \pi f C_0 \varepsilon_{r} (f)  } \quad ,
\label{eq:Z(f)_epsilon(f)}
\end{eqnarray} 
 with the geometrical capacitance $C_0$.
For a parallel-plate capacitor, $C_0 = \frac{\varepsilon_0 A}{d}$ with the vacuum permittivity $\varepsilon_0$, electrode area $A$ and plate distance $d$.
Separating real and imaginary parts in Eq.~(\ref{eq:Z(f)_epsilon(f)}), yields
\begin{eqnarray} 
Z(f)& = & \frac{\varepsilon_{r}'' (f)-i\varepsilon_{r}' (f)}{2 \pi f C_0 \left\{\varepsilon_{r}' (f)^2 + \varepsilon_{r}'' (f)^2 \right\}} \quad .
\label{eq:Z_epsilon}
\end{eqnarray} 
By inserting the logarithmic frequency dependence  Eq.~(\ref{eq:epsilon(f)}) into  Eq.~(\ref{eq:Z_epsilon}), we obtain the equivalent-circuit element $Z_{\rm{DW}}$,  which models the impedance of a ferroelectric due to pinning of domain walls in a potential valley [c.f.~Eq.~(\ref{eq:U_random})]  of a random energy landscape 

\begin{eqnarray} 
Z_{\rm{DW}} (f)& = & \frac{\varepsilon_{r,0}''-i\left[ \varepsilon_{r,0}'+ \Delta \varepsilon_{r}' \ln\left(\frac{1\,{\rm Hz}}{2 \pi f}\right)\right]}{2 \pi f C_0 \left\{\left[ \varepsilon_{r,0}'+ \Delta \varepsilon_{r}' \ln\left(\frac{1\,{\rm Hz}}{2 \pi f}\right)\right]^2 + \varepsilon_{r,0}''^2 \right\}} ,\nonumber\\
&&
\label{eq:Z_DW}
\end{eqnarray} 
with $\varepsilon_{r,0}'$, $\varepsilon_{r,0}''$ and $\Delta \varepsilon_{r}'$ as free parameters.
Consequently, $Z_{\rm{DW}}$ will be called the "domain wall pinning element" in the following.

Impedance spectroscopy allows one to extract the free parameters $\varepsilon_{r,0}'$, $\varepsilon_{r,0}''$ and $\Delta \varepsilon_{r}'$ by fitting the measured impedance over the full frequency range to the impedance of an appropriate equivalent-circuit model of the capacitor stack, including the ferroelectric.
Since domain wall pinning occurs solely in the ferroelectric layer and not in the electrodes and interfaces, the impedance response of electrodes and interfaces do not exhibit the characteristic logarithmic frequency dispersion over the measured frequency range and hence can be discriminated by their different frequency dispersion from the domain wall pinning element which models the ferroelectric layer.
Thus, the different contributions (from electrodes, interfaces and the ferroelectric layer) to the measured impedance of the capacitor stack are deconvoluted, by equivalent-circuit fitting with the domain wall pinning element.

Measurements at different excitation field amplitudes $E_{0}$ then allow one to determine the field dependence of the dispersion parameters $\varepsilon_{r,0}'(E_0)$, $\varepsilon_{r,0}''(E_0)$ and $\Delta \varepsilon_{r}'(E_0)$.
Subsequently, the field- and frequency-dependent dielectric response $\varepsilon_r(E_0,f)$ of the ferroelectric is obtained by inserting the field-dependent dispersion parameters into Eq.~(\ref{eq:epsilon(f)}).
For comparison, the conventional approach for the analysis of complex relative permittivity measurements is described in appendix A.

Note that for systems in which the assumption of a broad distribution of relaxation times and/or $\Theta \sim 1$ is not fulfilled, similar equivalent-circuit elements can be obtained by inserting the corresponding frequency dispersion into Eq.~(\ref{eq:Z_epsilon}).

\section{Experimental} 
\label{sec:Experimental}

A ceramic BCZT target was used to grow 200-nm-thick epitaxial single crystalline films on (001)-oriented Nb-doped (0.5 wt\%) SrTiO$_{3}$ (Nb:STO) single crystal by means of pulsed laser deposition (PLD).
We used a 248\,nm KrF excimer laser with an energy of 110\,mJ and 5\,Hz repetition rate.
For the deposition on Nb:STO substrates, the temperature was fixed at $700\,^{\circ}$C with an oxygen partial pressure $p_{\rm O_2}=0.1\,$mbar.
After deposition, the films were heated up to $800\,^{\circ}$C and kept there for 15 minutes before cooling
down to $700\,^{\circ}$C.
Subsequently, $p_{\rm O_2}$ was increased to 5\,mbar and the films were cooled down to room temperature with a rate of 10\,K/min.

The microstructure of the films was investigated by x-ray diffraction (XRD) to obtain the crystallographic orientation and to check for possible secondary phases.
One sample was further analyzed by scanning transmission electron microscopy (STEM) to investigate possible crystal defects in BCZT, which may act as domain wall pinning-centers.

For electrical measurements, 3\,-mm diameter disc-shaped Au electrodes ($\sim\,$100-nm-thick) were deposited on top of the BCZT films by electron beam evaporation through a shadow mask.
Impedance spectra of the capacitor stack from $f=$ 10\,Hz to 1\,MHz were collected at different excitation field amplitudes $E_0$ using a Solartron 1260 impedance analyzer together with a probe station.
For each measurement, $E_0$  was successively increased  from 2.5 to  25\,kV/cm (applied root mean square voltage $V_{\text{RMS}}=0.05-0.5$\,V) with a step-size of 1.25\,kV/cm, which results in the measurement of the impedance under 19 different  $E_0$.

\section{Results and Discussion} 
\label{sec:Results}

In this section, we present XRD data and results from electrical measurements from one representative epitaxial BCZT device. The STEM results were obtained from a different epitaxial BCZT device which was fabricated with the same process parameters (see Sec. \ref{sec:Experimental}).

\subsection{XRD and STEM}
\label{subsec:XRD-STEM}

\begin{figure}[b] 
\includegraphics[width=1\columnwidth]{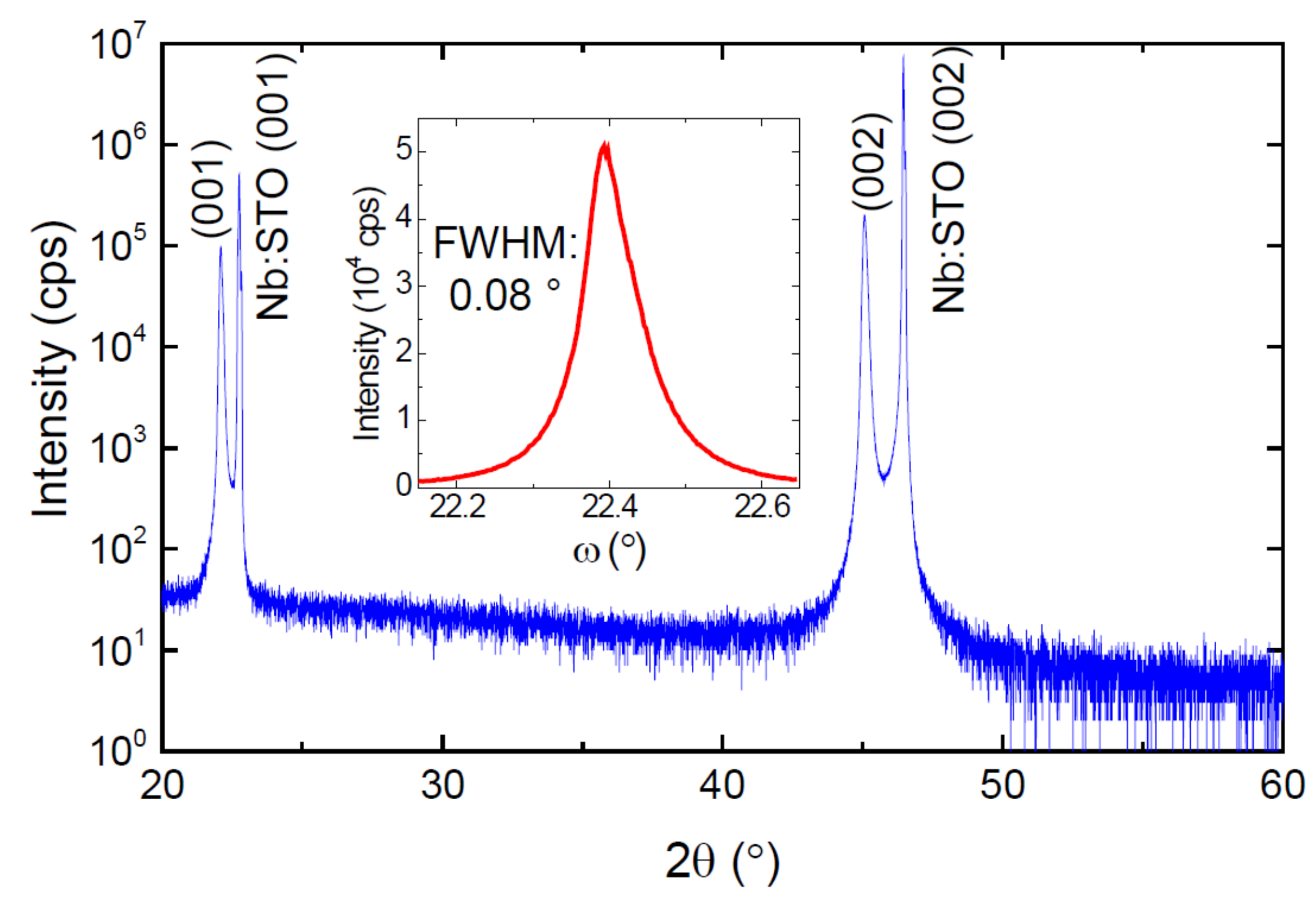}
\caption{XRD $\theta-2\theta$ scan of BCZT thin film on Nb:STO.
The inset shows the rocking curve of the (002) BCZT peak.}
\label{fig:XRD} 
\end{figure} 

Figure \ref{fig:XRD} shows the XRD $\theta-2\theta$ diffraction pattern for a BCZT thin film deposited on a (001)-oriented Nb:SrTiO$_3$ (Nb:STO) single crystal.
No traces of secondary phases are observable.
The structural phase diagram of BCZT shows four possible crystallographic phases, i.e.~cubic, tetragonal, orthorhombic and rhombohedral \cite{Keeble13}.
Due to the small variations in the lattice parameters of the possible phases, a rigorous specification of the crystallographic phase of the films is not accessible from the XRD data.
Thus, we denote the film structure as pseudo-cubic, which represents the four possible phases. For simplicity, we use cubic metrics for the Miller indices and crystal directions.
The film grown on a Nb:STO substrate is epitaxial and the full width half maximum (FWHM) of the rocking curve of the (002)-BCZT peak (inset in Figure \ref{fig:XRD}) has a value of $\omega=0.08\,^{\circ}$, indicating high crystalline quality of the film. The results are in good agreement with BCZT thin films grown by PLD on undoped STO \cite{Scarisoreanu15} and SrRuO$_{3}$ coated STO \cite{Lin15}.

\begin{figure}[t] 
\includegraphics[width=\columnwidth]{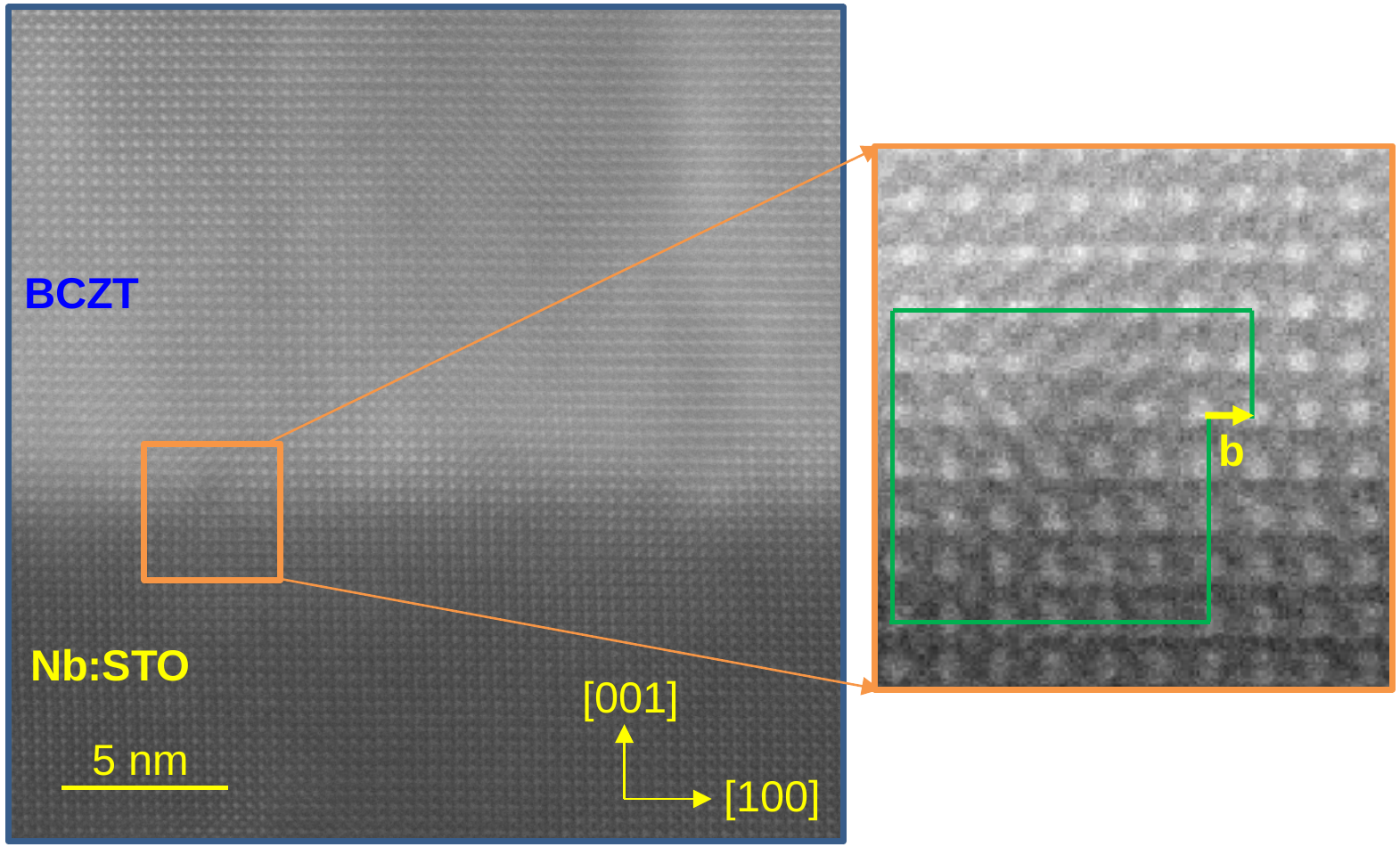}
\caption{Cross-section STEM image of Nb:STO/BCZT interface viewed along the {[}010{]} zone axis.
The enlarged section shows an edge-type misfit dislocation with Burgers vector $\bm{b}=a[100]$.
}
\label{fig:STEM} 
\end{figure} 

The epitaxial growth  of BCZT on Nb:STO was also confirmed by STEM.
Moreover, our STEM analysis revealed the occurrence of edge-type misfit dislocations at the epitaxial Nb:STO/BCZT interface, characterized by the Burgers vector $\bm{b}=a[100]$.
Here, $a$ denotes the lattice constant along the [100] direction within the BCZT unit cell (Fig.~\ref{fig:STEM}).
These misfit dislocations act as possible domain wall pinning centers, which was already observed in similar epitaxial PZT heterostructures \cite{Su11}.

\subsection{Impedance spectroscopy and equivalent-circuit analysis}
\label{subsec:Z-spectroscopy}

\begin{figure}[t] 
\includegraphics[width=1\columnwidth]{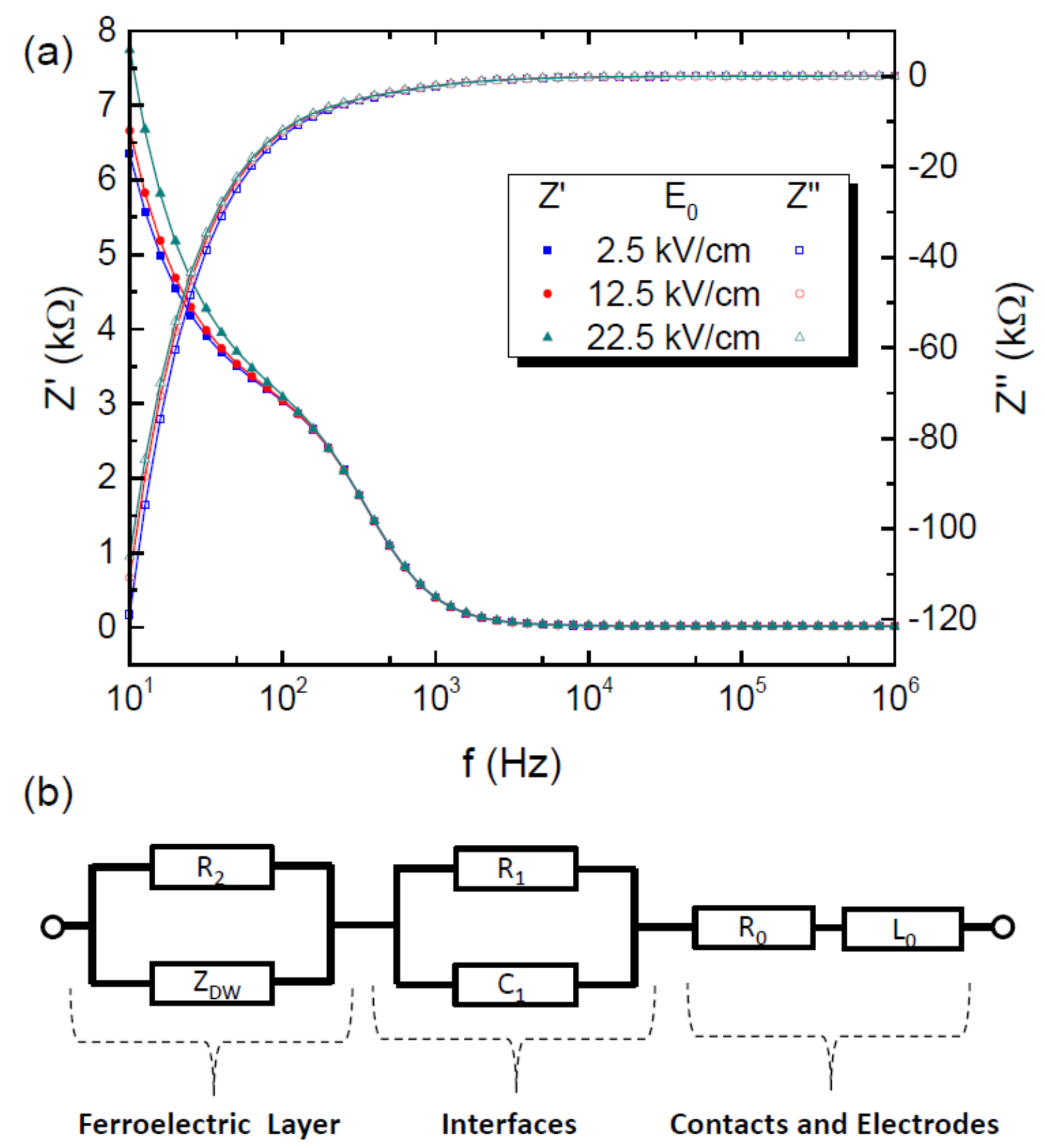} 
\caption{(a) Bode plots $Z'(f)$ and $Z''(f)$ obtained by impedance spectroscopy (symbols) for a selection of different $E_0$ on epitaxial BCZT thin film capacitor.
The equivalent circuit in (b) was used to fit the data (solid lines).}
\label{fig:Bode} 
\end{figure} 

\begin{table*}[t] 
\caption{Fit parameters for the epitaxial BCZT capacitor stack at three different excitation field amplitudes $E_0$. The estimated standard errors for $R_2$ are below 1 k$\Omega$.
}
\begin{ruledtabular}
\begin{tabular}{cccccccccc}
$E_0$ (kV/cm)	&$R_0$ ($\Omega$)		&$R_1$ ($\Omega$)	           & $C_1$ (nF)	          &$ R_2$ (M$\Omega$)          &$\varepsilon_{r,0}'$	 &$\Delta \varepsilon_{r}'$      & $\varepsilon_{r,0}''$  &($\mu$H) \\ \hline 
2.5			& $18\pm 1$ 			& $2889\pm 4$			& $150.5\pm 0.3$	 & $29$ 	            & $457.9\pm 0.2$		& $7.27\pm 0.04$                    &$10.65\pm 0.03$   &$4.8\pm 0.8$ \\
12.5		& $16\pm 1$			      & $2880\pm 3$		      & $151.8\pm 0.2$	 & $21$ 	            & $493.7\pm 0.1$ 		& $8.41\pm 0.03$                    &$13.23\pm 0.02$    &$5.3\pm 0.5$ \\
22.5	      & $14\pm 3$ 			      & $2838\pm 7$		      & $155.1\pm 0.6$ 	 & $13$ 	            & $526.4\pm 0.4$	 	& $11.41\pm 0.09$                   &$18.23\pm 0.07$   &$5.3\pm 1.4$
\end{tabular}
\end{ruledtabular}
\label{Tab-1}
\end{table*} 

Collected impedance spectroscopy data (for a selection of three different $E_0$) on the epitaxial device are shown in Fig.~\ref{fig:Bode}(a).
A validation procedure to rule out artifacts in the measured impedance spectra is given in appendix B.
The equivalent circuit depicted in Fig.~\ref{fig:Bode}(b) was used to fit the collected impedance spectra \cite{mathematica_script}.
Here the series resistance $R_{0}$ corresponds to the resistance of the cables,  measurement probes and the electrodes. The $R_{1}C_{1}$-Element represents the electrode-BCZT interfaces.
The resistance $R_{2}$  and $Z_{\text{DW}}$ model the ferroelectric film inside the capacitor stack.
$R_{2}$  is attributed to losses due to possible leakage currents, whereas the losses due to domain wall motion are contained in $Z_{\text{DW}}$.
Additionally, the dispersion parameters are contained in the domain wall pinning element.
The element $L_{0}$ models the inductance of the cables used.
Results of the fits can be seen in the Bode plot \cite{Macdonald18} shown in Fig.~\ref{fig:Bode}(a) where the real and the imaginary part of the impedance $Z=Z'-iZ''$ are plotted vs frequency (for three values of $E_0$).
An additional logarithmic Bode plot of the fits is given in appendix C.
In all cases, the fits are  in good agreement with the measured data, indicating that the utilized equivalent circuit is a suitable model for the thin film capacitors.
Fit parameters obtained from equivalent-circuit fits at three different excitation field amplitudes are summarized in Table \ref{Tab-1}.
The standard errors are calculated from maximum likelihood estimation \cite{Hoel1962}.

\begin{figure}[t] 
\includegraphics[width=1\columnwidth]{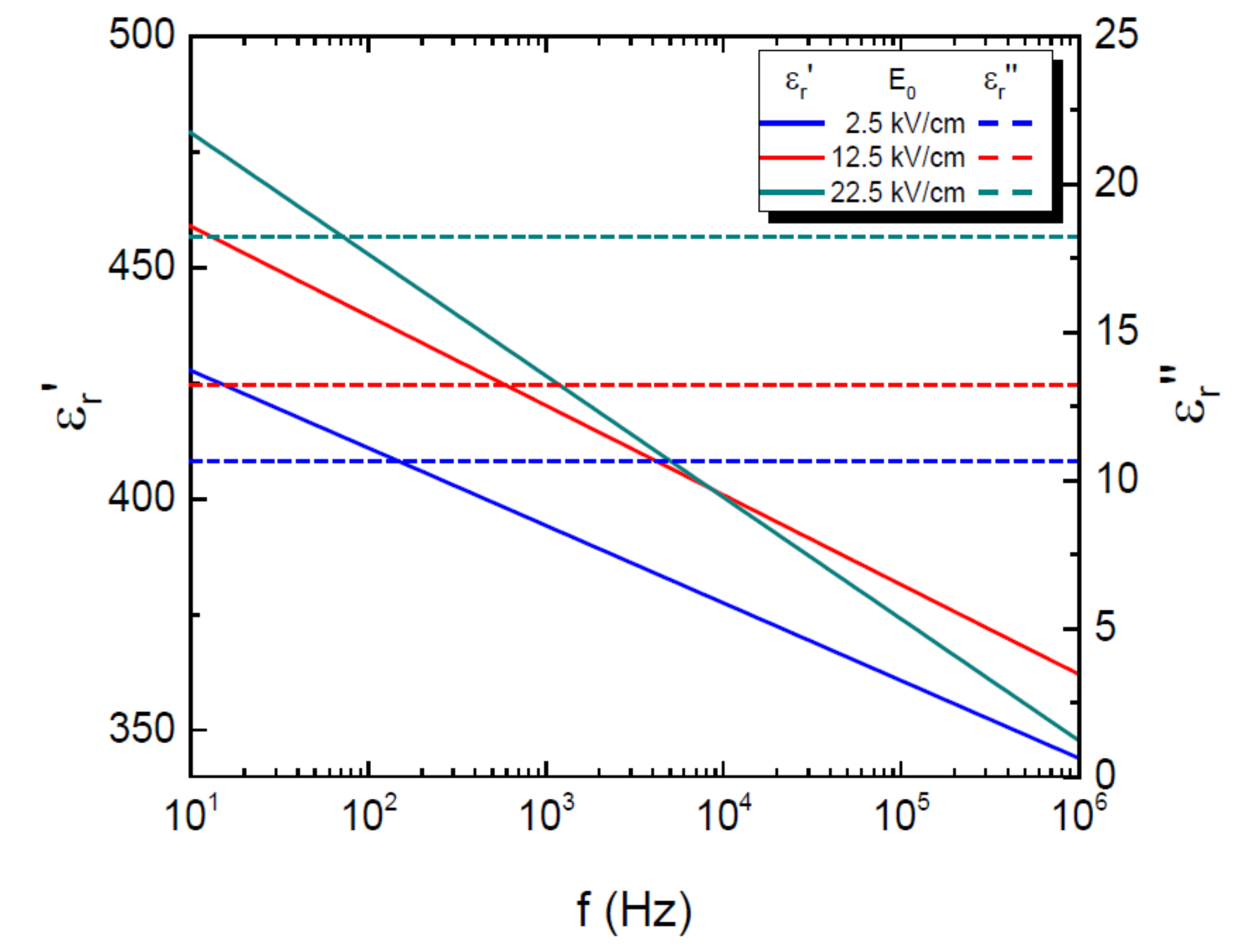}
\caption{Complex relative permittivity $\varepsilon_{r}(f)$ of the BCZT film for a selection of three different excitation field amplitudes $E_0$.}
\label{fig:epsilon_f} 
\end{figure} 

Subsequently, the complex relative permittivity $\varepsilon_{r}(f)$ of the BCZT film (i.e., without contributions from the electrodes and ferroelectric/electrode interfaces) was calculated from the obtained fit parameters according to Eq.~(\ref{eq:epsilon(f)}) for each field amplitude $E_{0}$.
The results for a selection of three different excitation field amplitudes are depicted in Fig.~\ref{fig:epsilon_f}.
We clearly observe  [c.f.~Fig.~\ref{fig:epsilon_f} and Table~\ref{Tab-1}] that $\varepsilon_{r}(f)$  and the parameters $\varepsilon_{r,0}'$, $\varepsilon_{r,0}''$ and $\Delta \varepsilon_{r}'$  depend on the field amplitude $E_0$, which indicates that dielectric nonlinearity and frequency dispersion are coupled in the BCZT films.

\subsection{Extended Rayleigh analysis}
\label{subsec:data-analysis}

\begin{figure}[b] 
\includegraphics[width=1\columnwidth]{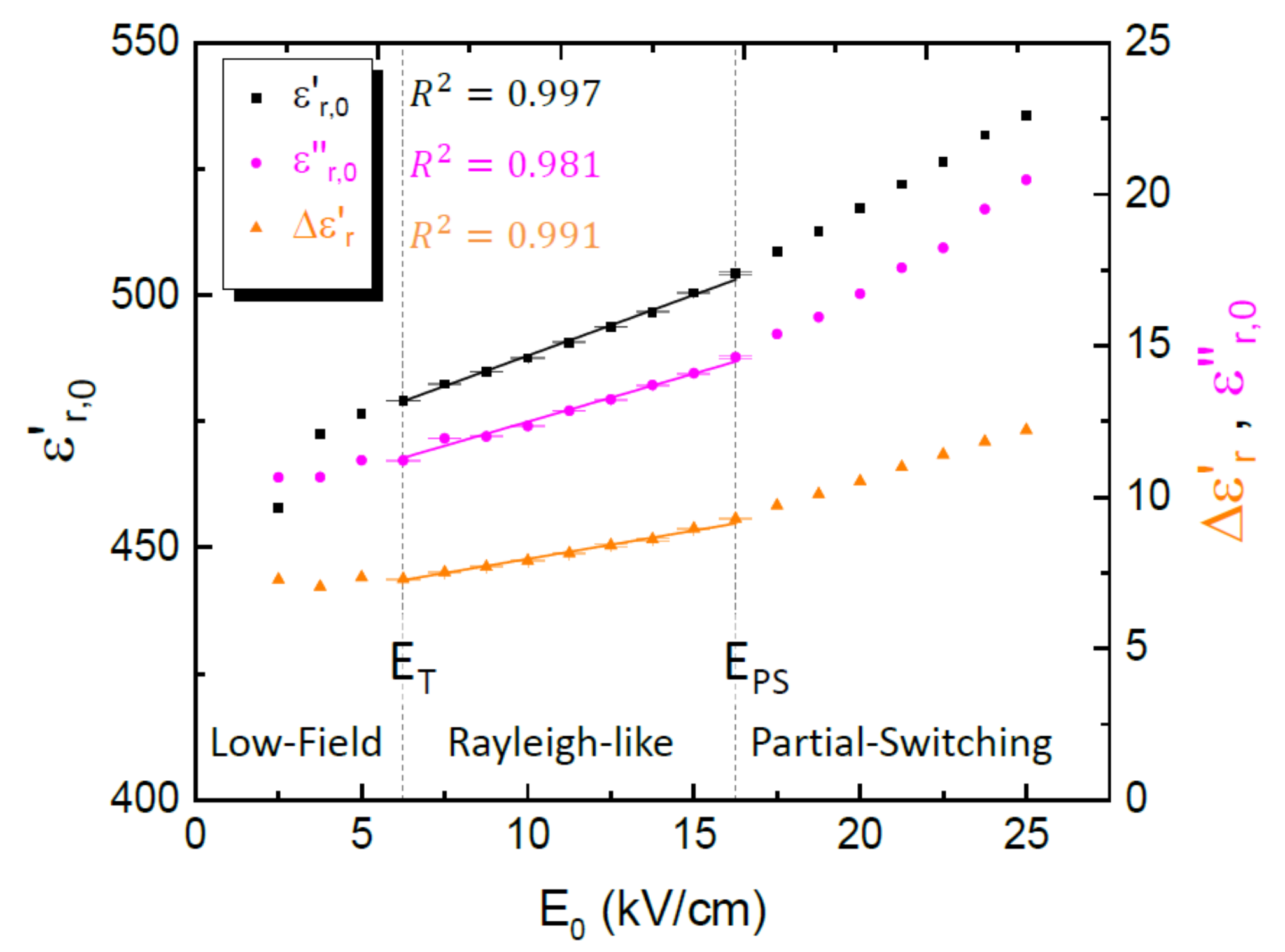}
\caption{Field dependence of the dispersion parameters $\varepsilon_{r,0}'(E_0)$, $\varepsilon_{r,0}''(E_0)$ and $\Delta \varepsilon_{r}'(E_0)$ of the BCZT film. The solid lines are linear fits in the region between the two vertical dashed lines,
which mark the threshold field $E_{\rm{T}}$ and the partial switching field $E_{\rm{PS}}$.}
\label{fig:DispersionParameters} 
\end{figure}

From the equivalent-circuit fits, the field dependence of the dispersion parameters $\varepsilon_{r,0}'(E_0)$, $\varepsilon_{r,0}''(E_0)$ and $\Delta \varepsilon_{r}'(E_0)$ can be extracted and the results are depicted in Fig.~\ref{fig:DispersionParameters}.

We observe that all dispersion parameters exhibit a linear dependence on the field amplitude $E_0$, i.e. in the guise of the following linear equations 
\begin{eqnarray} 
\varepsilon_{r,0}' (E_{0})& = & a_{0}'+a'E_{0}\quad ,   \nonumber \\
\varepsilon_{r,0}'' (E_{0})& =& a_{0}''+a''E_{0} \quad , \nonumber \\
\Delta \varepsilon_{r}' (E_{0})& = & b_{0}'+b'E_{0} \quad .
\label{eq:dispersion(E_0)}
\end{eqnarray} 
To specify the coefficients in  Eq.~(\ref{eq:dispersion(E_0)}), we insert  Eq.~(\ref{eq:dispersion(E_0)}) into  Eq.~(\ref{eq:epsilon(f)}), which yields
\begin{eqnarray} 
\varepsilon_{r}'(E_{0},f)& =  & [a_{0}'+a'E_{0}]+[b_{0}'+b'E_{0}]\ln\left(\frac{1\,{\rm Hz}}{2 \pi f}\right)  \quad ,  \nonumber \\
\varepsilon_{r}''(E_{0},f)& = &[a_{0}''+a''E_{0}] \quad .
\label{eq:epsilon_general}
\end{eqnarray} 
Comparing the real part $\varepsilon_{r}'(E_{0},f)$ of Eq.~(\ref{eq:epsilon_general})  with  Eqs.~(\ref{eq:RayleighParameter}) and ~(\ref{eq:epsilon(E_0,f)}), gives rise to the identifications $a_{0}' \equiv \varepsilon_{r,\text{in},0}'$, $b_{0}' \equiv \Delta  \varepsilon_{r,\text{in}}'$, $a' \equiv \alpha_{0}'$ and $b' \equiv \Delta \alpha '$.
It should be noted, that similar identifications were previously demonstrated to be valid for a PZT-based ferroelectric ceramic \cite{Damjanovic1997}. 
However, the imaginary part of the material response was not considered in Ref.~\onlinecite{Damjanovic1997}. 
Here, comparison of the frequency-independent  imaginary part in Eq.~(\ref{eq:epsilon_general}), with the imaginary part of the Rayleigh law Eq.~(\ref{eq:epsilon(E_0)}) yields the additional identifications $a_{0}'' \equiv  \varepsilon_{r,\text{in}}''$ and $a'' \equiv \alpha''$.

Altogether, the field dependence of the dispersion parameters Eq.~(\ref{eq:dispersion(E_0)}) can then be written in the following form
\begin{eqnarray} 
\varepsilon_{r,0}' (E_{0})& = & \varepsilon_{r,\text{in},0}'+\alpha_{0}'E_{0}\quad ,   \nonumber \\
\varepsilon_{r,0}'' (E_{0})& =& \varepsilon_{r,\text{in}}''+ \alpha''E_{0} \quad , \nonumber \\
\Delta \varepsilon_{r}' (E_{0})& = &  \Delta  \varepsilon_{r,\text{in}}'+ \Delta \alpha 'E_{0} \quad ,
\label{eq:dispersion(E_0)_2}
\end{eqnarray} 
where the coefficient $\Delta \varepsilon_{r,\text{in}}'$ is the zero-field contribution to the logarithmic dispersion strength  and the field-dependent contribution is given by  $\Delta \alpha 'E_{0}$.

\begin{table*}[t!] 
\caption{Extracted parameters from the linear fits according to Eq.~(\ref{eq:dispersion(E_0)_2}). The estimated uncertainties from previous fitting steps were taken into account.
}
\begin{ruledtabular}
\begin{tabular}{cccccccccc}

	&$\varepsilon_{r,\text{in},0}'$	    &$\alpha_{0}'$  (cm/kV)	         & $\varepsilon_{r,\text{in}}''$	        &$ \alpha''$ (cm/kV)          &$\Delta  \varepsilon_{r,\text{in}}'$	       &$ \Delta \alpha '$ (cm/kV)    \\ \hline 

	& $463.9 \pm 0.1$ 			    & $2.41 \pm 0.01$			           & $9.31 \pm 0.02$	                & $0.318 \pm 0.002$ 	                            & $6.09 \pm 0.03$	        & $0.187 \pm 0.003$ 	                              
\end{tabular}
\end{ruledtabular}
\label{Tab-2}
\end{table*} 

Subsequently, linear fits according to Eq.~(\ref{eq:dispersion(E_0)_2}) were carried out  in the range $6.25\,\text{kV/cm} \leq E_{0} \leq 16.25\,\text{kV/cm}$ as indicated by the vertical dashed lines in Fig.~\ref{fig:DispersionParameters}.
The starting point for the linear fits at $E_0=6.25\,\text{kV/cm}\equiv E_{\rm{T}}$  was determined by the criterion that all subsequent data points are required to have a higher value compared to their left neighbor.
The endpoint for the linear fits was then chosen by the longest fitting range for which every linear fit has a correlation factor $R^2 \geq 0.980$, which is fulfilled for  $E_0=16.25\,\text{kV/cm} \equiv E_{\rm{PS}}$.
At higher fields, all three curves in Fig.~\ref{fig:DispersionParameters}, show an upward curvature which we attribute to the partial switching of domains \cite{Hall99}.
For the linear fits, the estimated standard errors from the previous equivalent-circuit fitting procedure were taken into account, which is also indicated by the error bars within the fitting range [c.f.~Fig.~\ref{fig:DispersionParameters}].
The results of the fits according to Eq.~(\ref{eq:dispersion(E_0)_2})  are summarized in Table \ref{Tab-2}.

\begin{figure}[b] 
\includegraphics[width=1\columnwidth]{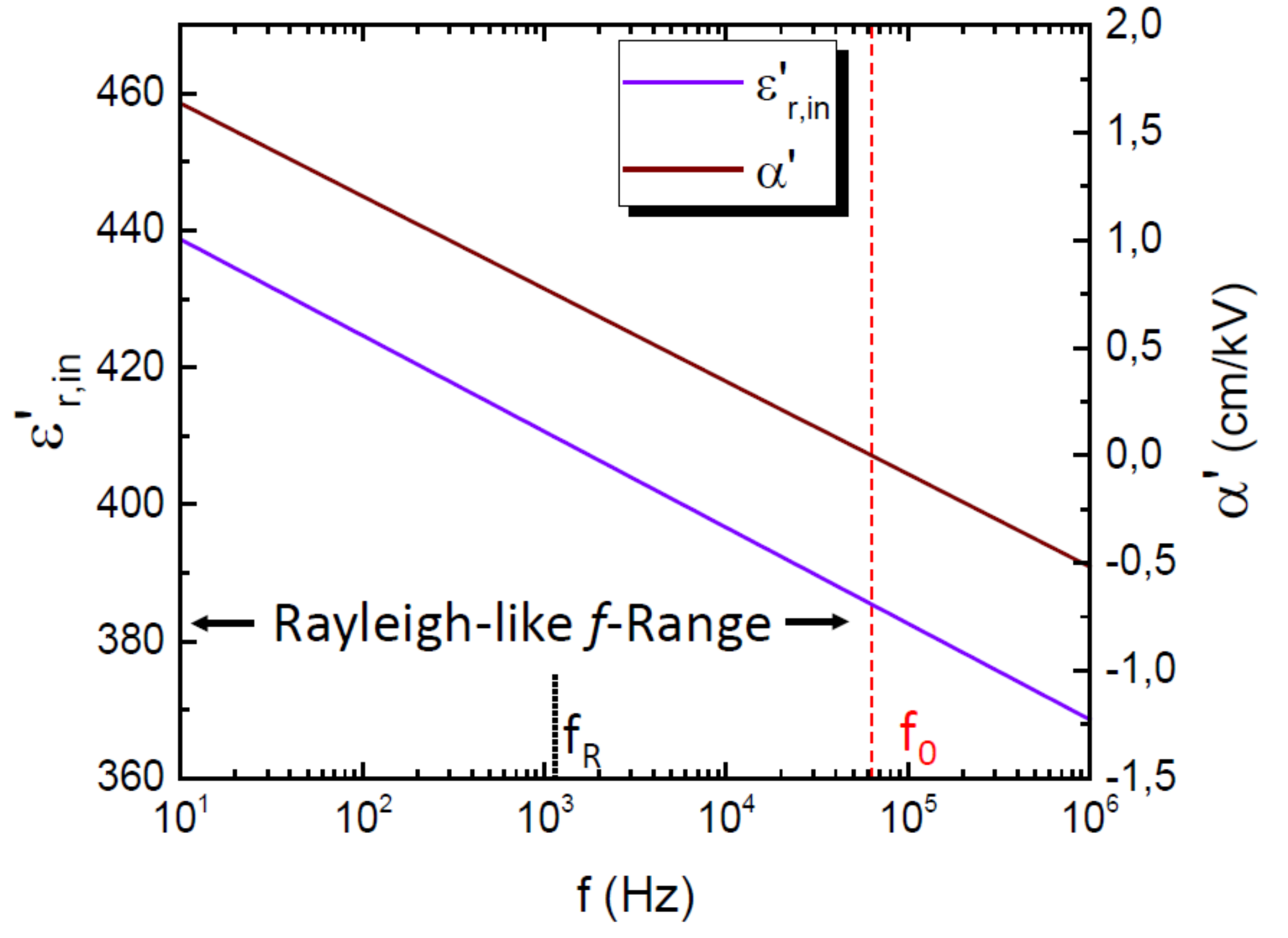}
\caption{Logarithmic frequency dispersion of reversible $\varepsilon_{r,\text{in}}'(f)$ and irreversible $\alpha'(f)$ Rayleigh parameters of the BCZT film.
The vertical dotted line at $\sim 1$ kHz marks the Rayleigh frequency $f_{\rm{R}}$.
The irreversible Rayleigh parameter changes to negative values at $f_0 \sim 63$ kHz, which is indicated by the vertical dashed line.}
\label{fig:RayleighParameters} 
\end{figure}

From the extracted fit parameters, the functional form of the frequency-dependent reversible and irreversible Rayleigh parameter of the BCZT thin film can be immediately obtained according to Eq.~(\ref{eq:RayleighParameter}) and the results are depicted in  Fig.~\ref{fig:RayleighParameters}.
Here, the vertical dotted line at$f_\text{R}\sim 1$ kHz indicates the Rayleigh frequency for the BCZT thin film, which we calculated according to Eq.~(\ref{eq:fR}).
As discussed in Sec.~\ref{sec:Theory}, $f_\text{R}$ is part of the Rayleigh-like frequency range, where the center of mass of the domain walls can move irreversibly via hopping between potential minima corresponding to the stochastic regime.

The dashed line at $f_0 \sim 63$ kHz marks the region where the irreversible Rayleigh parameter $\alpha'$ is zero and changes sign to negative values. 
This might indicate a transition from the stochastic regime to a pinned regime, in which the center of mass motion of the domain walls cannot follow the fast driving field, and thus the domain walls are captured in a potential valley. \cite{Fedorenko04}. 
Similar experimental indications for the freezing of the  irreversible center of mass motion of the domain walls at higher frequencies were reported previously \cite{Griggio12}.

By inserting the functional form of the field dependent frequency dispersion parameters  [c.f.~Fig.~\ref{fig:DispersionParameters} and Table \ref{Tab-2}]  into Eq.~(\ref{eq:epsilon(f)}), we finally obtain the functional form of the field and frequency dependent dielectric response in the epitaxial BCZT thin film, which is given by 
\begin{eqnarray} 
\varepsilon_{r}'(E_{0},f)& = & \varepsilon_{r,\text{in},0}'+\alpha_{0}' E_0+[\Delta  \varepsilon_{r,\text{in}}' + \Delta \alpha ' E_0]\ln\left(\frac{1\,{\rm Hz}}{2 \pi f}\right)  \nonumber \\
\varepsilon_{r}''(E_{0},f)& = & \varepsilon_{r,\text{in}}'' + \alpha ''E_0  .
\label{eq:epsilonBCZT}
\end{eqnarray} 
Note that $\varepsilon_{r,\text{in}}''$ in Eq.~(\ref{eq:epsilonBCZT}) is not zero, which could be an indication for an internal bias or stress field \cite{Schenk18}.

The coupling between field dependence and frequency dispersion in the BCZT thin film arises explicitly in Eq.~(\ref{eq:epsilonBCZT}) as the product  of $E_0$ and the $f$-dependent term. 
Thus, this gives rise to the identifiation of the coefficient $ \Delta \alpha '$ as the coupling strength beetween dielectric nonlinearity and frequency dispersion, which is defined as the change of logarithmic frequency dispersion strength with field amplitude - or equivalent - as the change of the irreversible Rayleigh parameter with the logarithm of frequency, i.e. 
\begin{eqnarray} 
\Delta \alpha '& \equiv & \frac{\partial \alpha'(f)}{\partial \ln\left(\frac{1\,{\rm Hz}}{2 \pi f}\right)}   \nonumber \\
                                     & \equiv & \frac{\partial \Delta \varepsilon_{r}'(E_0)}{\partial E_0}  \quad,
\label{eq:kappa}
\end{eqnarray} 
which has a value of 0.187 cm/kV in the epitaxial BCZT thin film. 
In fact, the negative values of the irreversible Rayleigh parameter are a direct consequence of the coupling between frequency dispersion and dielectric nonlinearity in the BCZT thin film.

\begin{figure}[t] 
\includegraphics[width=1\columnwidth]{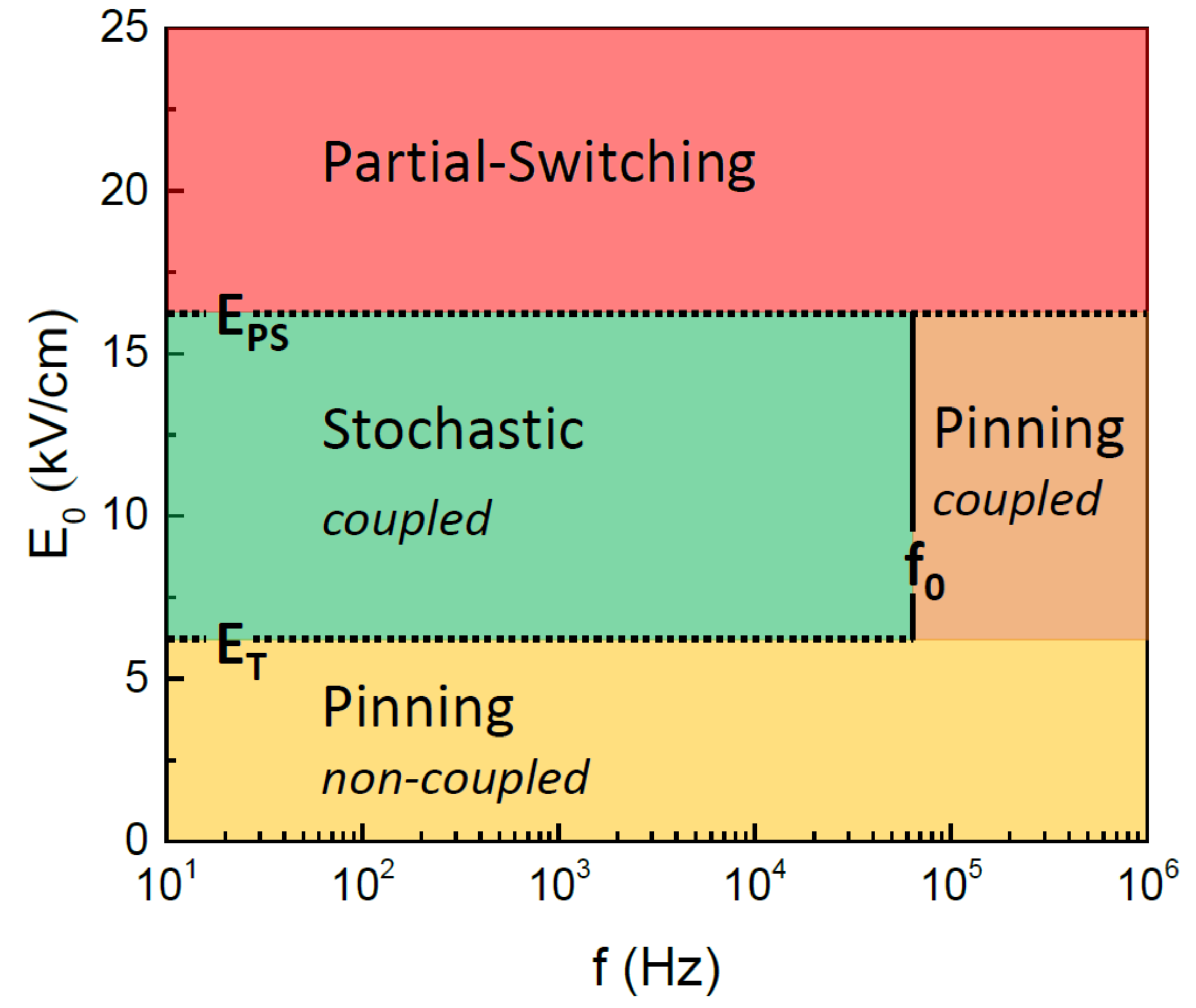}
\caption{Schematic diagram of the different domain wall motion regimes of the epitaxial BCZT thin film.
The threshold field $E_{\rm{T}}=6.25\,\text{kV/cm}$ and the partial switching field  $E_{\rm{PS}}=16.25\,\text{kV/cm}$ are obtained from the field dependence of the dispersion parameters [c. f.  Fig.~\ref{fig:DispersionParameters}] and  $f_0 \sim 63$ kHz is determined from the frequency dispersion of the irreversible Rayleigh parameter [c. f.  Fig.~\ref{fig:RayleighParameters}].}

\label{fig:PhaseDiagram} 
\end{figure} 

Altogether, our analysis yields the schematic diagram depicted in Fig.~\ref{fig:PhaseDiagram} for the different domain wall motion regimes in the epitaxial BCZT thin film.
Note, that the diagram constructed in Fig.~\ref{fig:PhaseDiagram} is not a generic diagram for any epitaxial BCZT film.
In fact, it is a schematic diagram of different domain-wall motion regimes, which can now be identified within the framework of the extended Rayleigh analysis.

At low field amplitudes $0 < E_0 <  E_{\rm T}$, the center of mass of the domain walls cannot overcome the potential well  Eq.~(\ref{eq:U_random}) of the random energy landscape, however, domain wall segments can jump between metastable states with close energies, which corresponds to the non-coupled pinning regime.
This observation has similarities to what has been found theoretically for disordered ferromagnets \cite{Nattermann01}.
By increasing the field amplitude above the threshold field $E_{\rm T}$, the center of mass of the domain walls can additionally jump between different potential minima, resulting in a coexistence of irreversible center of mass motion and relaxational motion of internal modes in the stochastic regime.
Note that this coexistence is not simply additive, due to the coupling of dielectric nonlinearity and frequency dispersion.
As a consequence of coupling, the irreversible Rayleigh parameter becomes zero at $f_0 \sim 63$ kHz and changes to negative values in the pinned coupling regime.
Note that in this coupled-pinning regime, domain wall segments can still jump between metastable states, which is indicated by the logarithmic frequency dispersion observed in this regime.
The  domain wall dynamics in the coupled-pinning  regime is not fully understood and remains as a task of a unified theory of dielectric nonlinearity and frequency dispersion, which is not yet available.

\section{Conclusion} 
\label{sec:Conclusion}

In conclusion, this work demonstrates several key aspects.
We introduce the domain wall pinning element $Z_{\rm{DW}}$  based on the theory of interface pining in random systems, which includes the characteristic logarithmic frequency dispersion due to domain wall pinning in a random energy landscape and hence models the impedance response of a ferroelectric below the coercive field.
Since domain wall pinning occurs solely in the ferroelectric and not in the electrodes and interfaces, the impedance response of electrodes and interfaces do not exhibit the characteristic logarithmic frequency dispersion and hence can be discriminated by their different frequency dispersion from the domain wall pinning element which models the ferroelectric.
Thus, different contributions to the measured overall impedance response are deconvoluted by equivalent-circuit fitting with the domain wall pinning element.
Moreover, we demonstrate the practical  application of this new element to extract the field- and frequency-dependent dielectric response in an epitaxial lead-free ferroelectric BCZT thin film embedded in a capacitor stack and we reveal its superiority over the conventional approach. 
In addition, we perform an extended Rayleigh analysis which results in the quantification of the coupling strength and the explicit functional form $\varepsilon_{r}(E_{0},f)$ of the coupled dielectric response in the BCZT film.
Finally, we present a schematic diagram of the different domain wall motion regimes in the BCZT film and discuss the corresponding domain wall dynamics.

The present work is intended to serve as a guideline for future work on ferroelectric materials, which includes to reveal the effect of temperature and phase transitions \cite{Piorra19}, external bias fields \cite{Bassiri-Gharb07}, dopants \cite{Yang19}, ion-bombardement \cite{Saremi18, Saremi19} and substrate clamping \cite{Griggio12} on domain wall dynamics and related material properties.
This  may provide new insights  into nonlinear contributions to the dielectric- and piezoelectric responses, which is also of practical importance for many microelectromechanical systems (MEMS)  and may help to develop a unified theory of frequency dispersion and dielectric nonlinearity in ferroelectrics.


\section*{Acknowledgements}
\label{sec:Acknowledgements}

We gratefully acknowledge fruitful discussions with C.~Warres (NMI), C.~Hofer  and J.~C.~Meyer and technical support by M.~Turad and R. L\"offler (LISA$^+$).
This work was partially funded by the Bundesministerium f\"ur Bildung und Forschung (BMBF) under Grant No. 13GW0123E and partly funded by the Europ\"aische Fonds f\"ur regionale Entwicklung (EFRE) under Grant No. 712303.


\section*{Appendix A: Conventional Approach}
\label{sec:appendixA}

\begin{figure}[b!] 
\includegraphics[width=1\columnwidth]{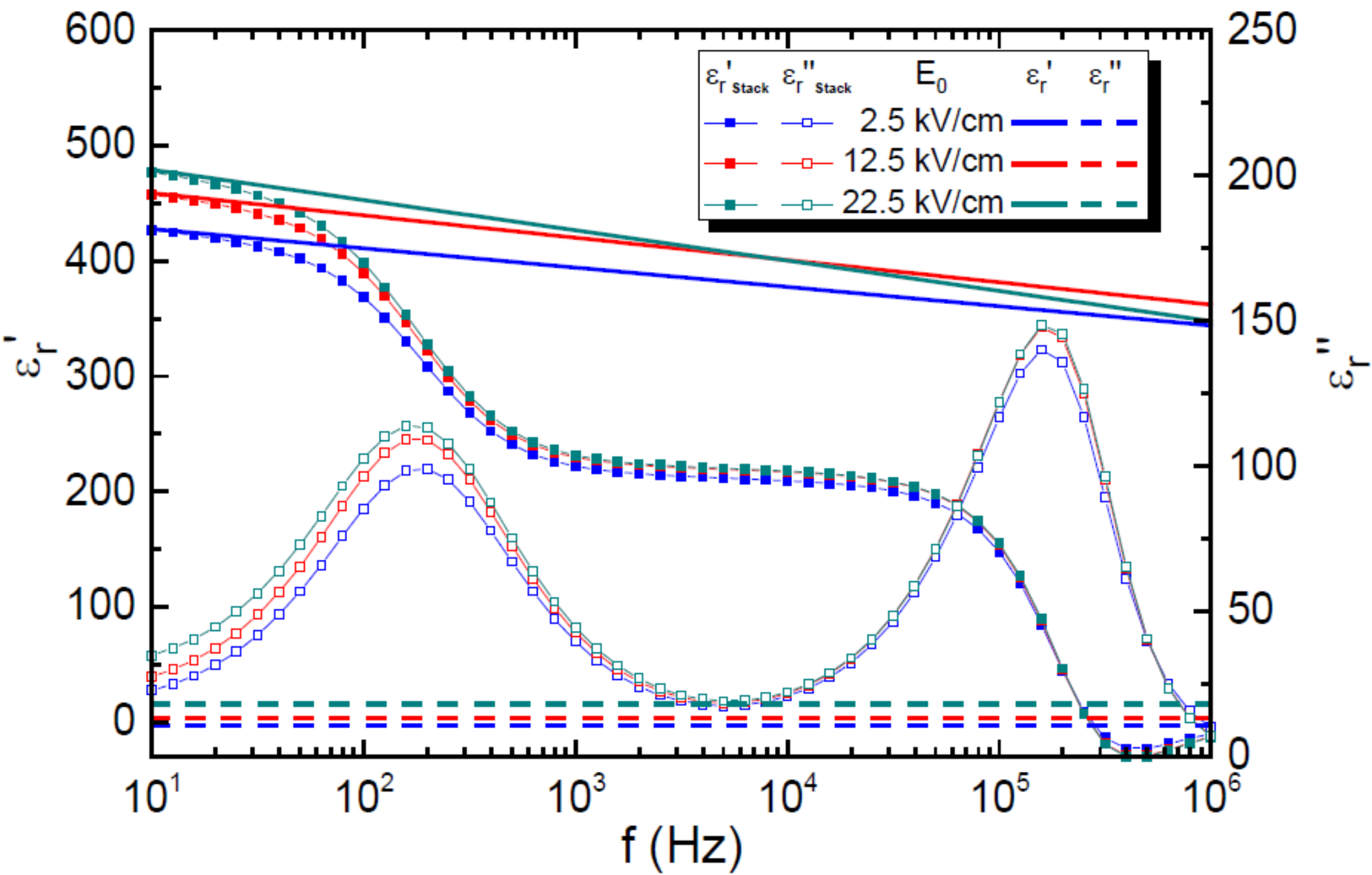}
\caption{Measured permittivities $\varepsilon_{r,\text{stack}}'(f)$ (full symbols) and $\varepsilon_{r,\text{stack}}''(f)$ (open symbols) of the BCZT capacitor stack for a selection of different excitation field amplitudes $E_0$.
For clarity, the symbols are connected by thin lines which do not represent fits to the data.
For comparison, the extracted permittivities of the BCZT layer from the equivalent-circuit fits [c.f.~Fig.~\ref{fig:epsilon_f} and Eq.~(\ref{eq:epsilon(f)})] are also shown (solid and dashed thick lines).
}
\label{fig:epsilon_stacks} 
\end{figure} 

For the conventional approach, the measured complex impedance $Z=Z'-iZ''$ of the capacitor stacks is converted into complex permittivity data of the capacitor stacks using the formalism described in Ref.~\onlinecite{Gerhardt94}, resulting in 
\begin{eqnarray} 
\varepsilon_{r,\text{stack}}'(f)&=&\frac{-Z''}{2\pi f\varepsilon_{0}(Z'^{2}+Z''^{2})}\frac{d}{A} \quad ,\nonumber \\
\varepsilon_{r,\text{stack}}''(f)&=&\frac{Z'}{2\pi f\varepsilon_{0}(Z'^{2}+Z''^{2})}\frac{d}{A} \quad .  
\label{eq:epsilon_stack} 
\end{eqnarray}
The resulting  $\varepsilon_{r,\text{stack}}(f)$ plots are shown in Figure \ref{fig:epsilon_stacks} for a selection of three different $E_0$.  For comparison, also the extracted complex permittivity $\varepsilon_{r}(f)$ of the BCZT layer from the equivalent-circuit fits [c.f.~Fig.~\ref{fig:epsilon_f} and Eq.~(\ref{eq:epsilon(f)})] is depicted.
The capacitor stack exhibits a Debye-like dielectric relaxation at high frequencies ($\sim 10^5\,$Hz), due to to resistive losses in the electrodes \cite{Tyunina06}.
Moreover, the dielectric spectrum of the epitaxial thin film capacitor indicates an additional Debye-like relaxation process at lower frequencies ($\sim 100\,$Hz).
We attribute this second relaxation process to an extrinsic electrode/film interface effect, which was previously reported to occur in other ferroelectric thin films grown on Nb:STO substrates \cite{Schmidt2012}.
This is also consistent with the theoretical discussion of the potential well  Eq.~(\ref{eq:U_random}) within a random energy landscape, which leads to a distribution of relaxation times in the BCZT layer and hence the BCZT layer is not expected to exhibit a Debye-like dielectric relaxation which corresponds to a single relaxation time and hence one dominant relaxation frequency $f_{\rm{r}}$.
The relaxation frequency can be calculated from the obtained interface fit parameters $R_1$ and $C_1$ according to the relation $f_{\rm{r}}=1/2\pi R_1 C_1$.
The calculated relaxation frequencies are in agreement with the peaks in the imaginary part of permittvity of the capacitor stack [c.f.~\ref{fig:epsilon_stacks}] at frequencies ($\sim 100\,$Hz), which indicates the correct deconvolution of film and interface contributions by our equivalent-circuit fits.
Moreover, for BCZT thin films grown on platinized Si substrates, the second Debye-like dielectric relaxation was absent \cite{BeckerJAP20}.
Furthermore, we note that the slightly negative values of  $\varepsilon_{r,\text{stack}}'(f)$ at frequencies around 1 MHz are due to the inductance of the cables used \cite{Macdonald18}.

It is clear, that the extraction of frequency dispersion of permittivity  from  [c.f.~Eq.~(\ref{eq:epsilon_stack}) and symbols in Fig.~\ref{fig:epsilon_stacks}] does not reflect the true dielectric dispersion of the BCZT thin film, since there is no deconvolution of electronically distinct components forming the capacitor heterostructure.
The same is true for the extraction of Rayleigh parameters, which has been recently demonstrated \cite{Schenk18}.

\section*{Appendix B: Validation of impedance spectra}
\label{sec:appendixB}

\begin{figure}[t] 
\includegraphics[width=1\columnwidth]{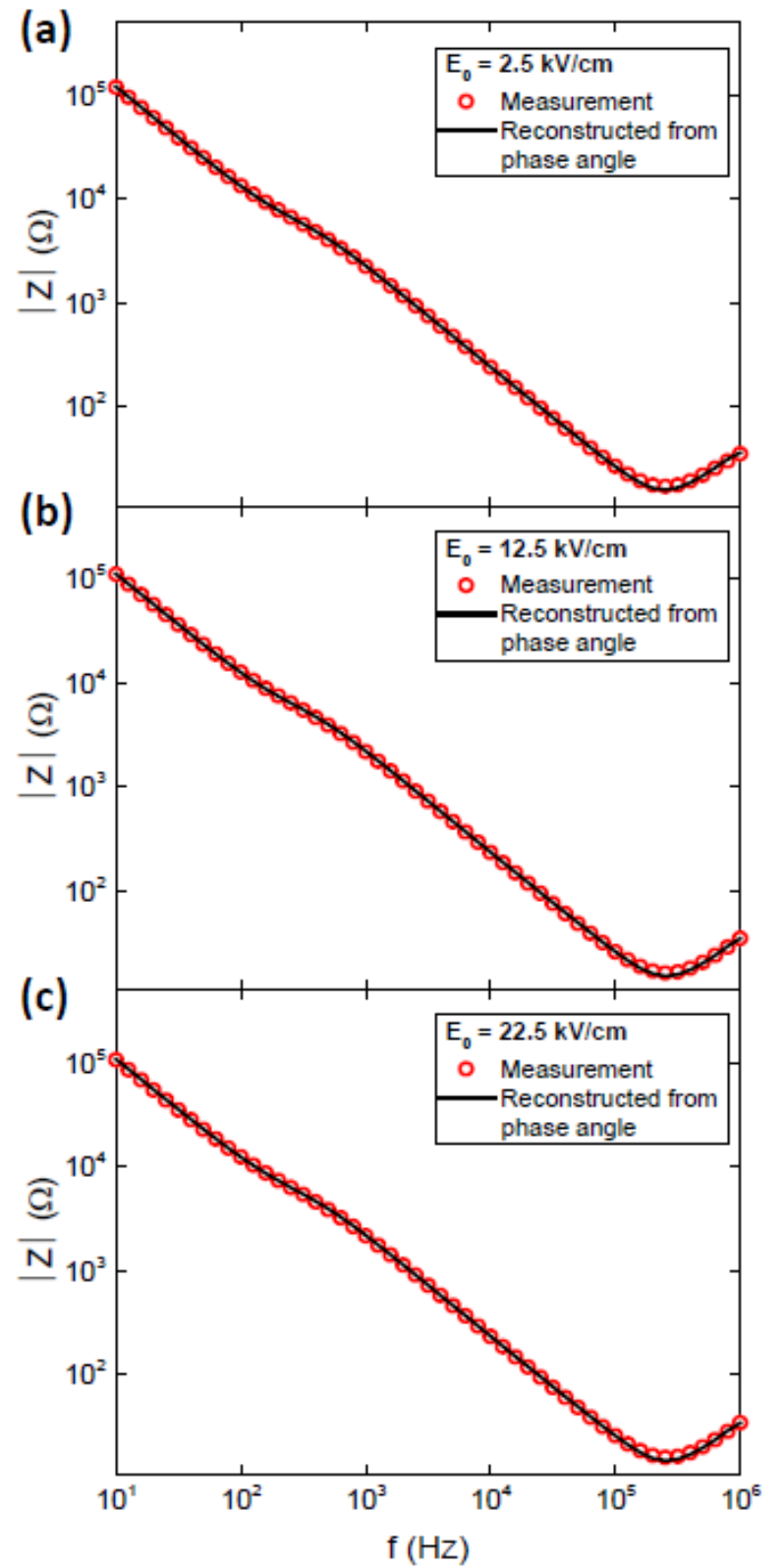}
\caption{Comparison of measured (symbols) and Z-HIT-reconstructed (black solid lines) impedance modulus $\vert Z \vert$ of the BCZT capacitor stack vs frequency $f$ for (a) $E_{0}$ = 2.5 kV/cm, (b) $E_{0}$ = 12.5 kV/cm and (c) $E_{0}$ = 22.5 kV/cm.
}
\label{fig:ZHIT} 
\end{figure}

Impedance spectra might be afflicted with artifacts.
To rule out artifacts introduced by experimental set-up conditions in the measured impedance spectra, we use the two-pole Hilbert transform (Z-HIT) described in Ref.~\onlinecite{Schiller01}.
This approach is beneficial compared to methods based on Kramers-Kronig (KK) relations \cite{Urquidi-Macdonald1990,Agarwal1995}, since the KK relations are strictly defined within the frequency range between zero and infinity and hence, the extrapolation of the measured impedance data is a unavoidable task which may lead to erroneous results \cite{Schiller01}.
The Z-HIT avoids the necessity of an extrapolation of the frequency range and allows one to reconstruct the modulus of the impedance $\vert Z \vert$ vs $f$ from the measured values of the less artifact-prone \cite{Schiller01,Schenk18} phase angle.
The results of the Z-HIT for three different excitation field amplitudes $E_0$ are depicted in Fig.~\ref{fig:ZHIT}.
In all cases, the reconstructed curves match fairly well with the measured data, which indicates that the measured impedance spectra are not afflicted with artifacts and the investigated system is stable, causal, (sufficiently) linear \cite{Schenk18} and exhibits finite values over the entire frequency range.

\section*{Appendix C: Logarithmic Bode Plot}
\label{sec:appendixC}

\begin{figure}[t] 
\includegraphics[width=1\columnwidth]{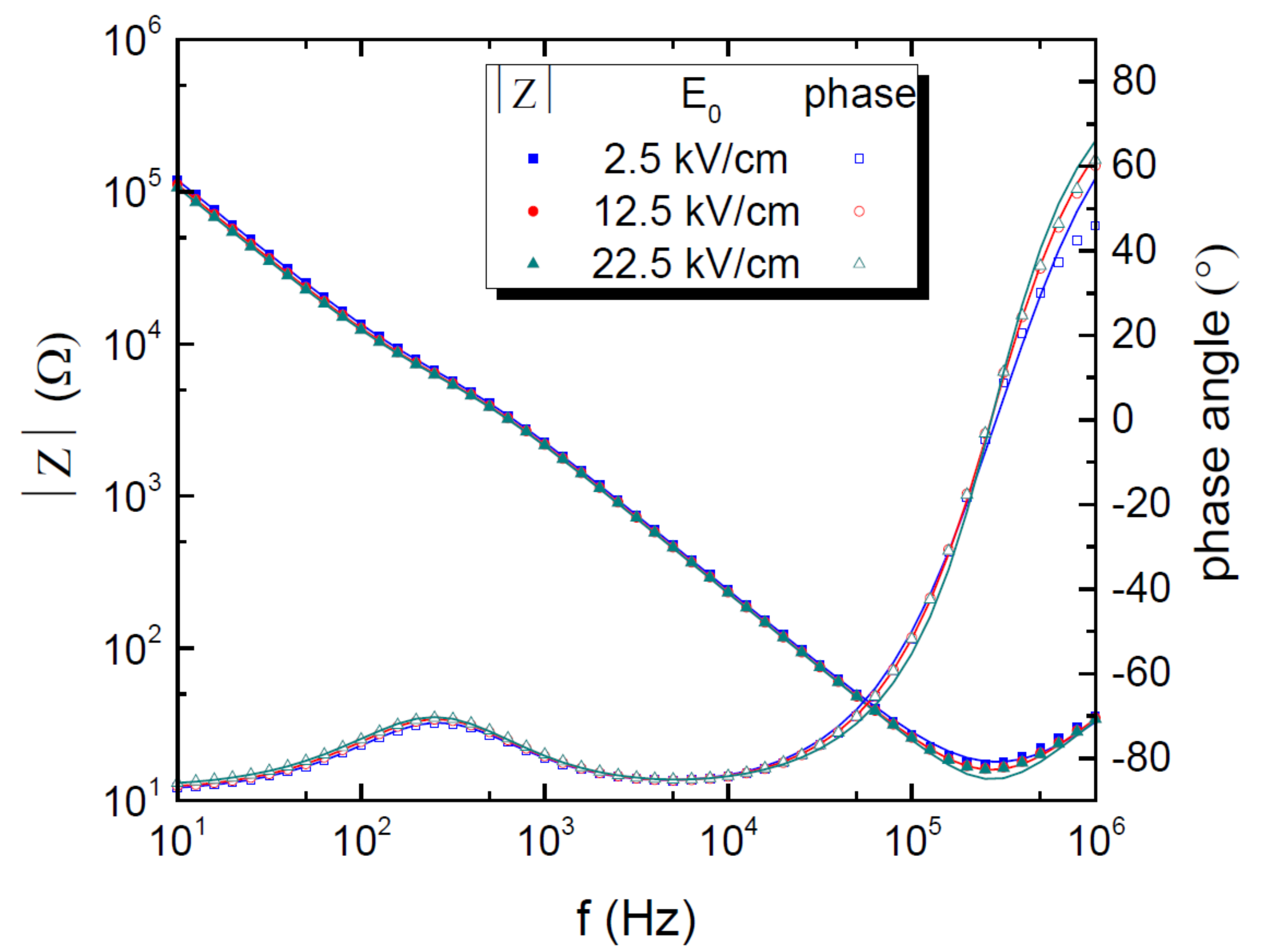}
\caption{Logarithmic Bode plot of impedance modulus $\vert Z \vert$ and phase angle obtained by impedance spectroscopy (symbols) and the corresponding equivalent-circuit fits (solid lines)  for a selection of different $E_0$ on the epitaxial BCZT thin-film capacitor stack.
}
\label{fig:zphase} 
\end{figure}

To further check the overall quality of the equivalent-circuit fits, the impedance modulus $\vert Z \vert$ and the phase angle are depicted in the logarithmic Bode plot in Fig.~\ref{fig:zphase}.
In all cases, the fits are in good agreement with the measured data which indicates the validity of the equivalent-circuit model  [c.f.~Fig.~\ref{fig:Bode}(b)].

\clearpage

\bibliography{DE-Response-BCZT}
\end{document}